\documentclass{appolb1}
\usepackage{graphicx}
\usepackage{amsmath}
\usepackage{slashed}
\usepackage{cite}
\usepackage[colorlinks=true,citecolor=blue]{hyperref}
\begin{document}
% \eqsec  
% uncomment this line to get equations numbered by (sec.num)

\title{{\small DESY 17--188, DO-TH 17/31, TTK-17-36, NIKHEF 2017-059. \hfill}\\
Heavy~quark~form~factors~at~two~loops~in~perturbative~QCD\thanks{Presented by N. Rana at XLI International Conference 
of Theoretical Physics ``Matter to the Deepest'', Podlesice, Poland, September 3-8, 2017 and RADCOR 2017, St.~Gilgen 
Austria, September 24-29, 2017.}% % 
}

\author{J.~Ablinger$^a$, A.~Behring$^b$, J.~Bl\"umlein$^c$, G.~Falcioni$^d$, A.~De~Freitas$^c$,
P.~Marquard$^c$, N.~Rana$^c$, C.~Schneider$^a$
\address{
$^a$~Research Institute for Symbolic Computation (RISC),\\ Johannes Kepler University, Altenbergerstra\ss{}e 69, A--4040, Linz, Austria
\\
$^b$~Institut f\"ur Theoretische Teilchenphysik und Kosmologie,\\ RWTH Aachen University, D-52056 Aachen, Germany
\\
$^c$~Deutsches Elektronen--Synchrotron, DESY,\\ Platanenallee 6, D-15738 Zeuthen, Germany
\\
$^d$~Nikhef Theory Group,\\ Science Park 105, 1098 XG Amsterdam, The Netherlands
}
}
\maketitle
\begin{abstract}
\noindent
\noindent
We present the results for heavy quark form factors at two-loop order in perturbative QCD for different currents, namely vector,
axial-vector, scalar and pseudo-scalar currents, up to second order in the dimensional regularization parameter.
We outline the necessary computational details, ultraviolet renormalization and corresponding universal infrared structure.
\end{abstract}
\PACS{PACS number}
  
%% definition here
% 
\def\ep{\varepsilon}
\def\asr{\left( \frac{\alpha_s}{4 \pi} \right)}
\def\b0{\beta_0}

\def\q{\slashed{q}}
\def\F{{\rm F}}

\section{Introduction}

The abundance of top quark pair production at high energy colliders provides important precision tests, with a
strong potential for beyond the Standard Model (BSM) physics scenarios. The top quark, as the heaviest 
particle of the SM, has not been explored at high precision yet. Hence, detailed studies of this channel at future 
linear or circular electron-positron colliders is a crucial topic, which is likewise the case for the LHC. In order 
to match the experimental accuracy, precise predictions are required on the theoretical side as well. 
Furthermore, the form factors involving heavy quarks play an important role 
in determining various physical quantities concerning top quark pair production.
The vector and axial vector 
massive form factors are important building blocks for the forward-backward asymmetry 
in the production of bottom or top quarks at electron-positron colliders.
The decay of a scalar or pseudo-scalar particle to a pair of heavy quarks also
could play a very important role in shedding light on the quantum nature of the Higgs boson.
There are also static quantities like the anomalous magnetic moment, which receive
contributions from such massive form factors. For these reasons, phenomenology and 
higher order Quantum chromodynamics (QCD) corrections to these form factors
have gained much attention during the last decade.

A plethora of works \cite{Braaten:1980yq, Sakai:1980fa, Inami:1980qp, Drees:1990dq,
Arbuzov:1991pr, Djouadi:1994wt,
Altarelli:1992fs, Ravindran:1998jw, Catani:1999nf,
Gorishnii:1983cu,Gorishnii:1991zr,Surguladze:1994em,Surguladze:1994gc,Larin:1995sq,Chetyrkin:1995pd,Harlander:1997xa,Harlander:2003ai}
was followed by 
% 
% In \cite{Arbuzov:1991pr, Djouadi:1994wt}, the first order QCD corrections were obtained for 
% the vector and axial-vector form factors. The leading terms from the next-to-next-to-leading order (NNLO) 
% QCD corrections was captured in \cite{Catani:1999nf} followed by \cite{Altarelli:1992fs, Ravindran:1998jw}.
% On the other hand, the next-to-leading order (NLO) contributions to the scalar and pseudo-scalar form factors
% were known \cite{Braaten:1980yq, Sakai:1980fa, Inami:1980qp, Drees:1990dq} for long and NNLO corrections 
% by employing quark mass expansion to various orders, by \cite{Gorishnii:1983cu,Gorishnii:1991zr,Surguladze:1994em,Surguladze:1994gc,Larin:1995sq,Chetyrkin:1995pd,Harlander:1997xa,Harlander:2003ai}. 
a series of papers obtaining the two-loop QCD corrections for the vector form factor \cite{Bernreuther:2004ih}, 
the axial-vector form factor \cite{Bernreuther:2004th}, the anomaly contributions \cite{Bernreuther:2005rw}
and the scalar and pseudo-scalar form factors \cite{Bernreuther:2005gw}. 
An independent cross-check of the vector form factor has been performed in \cite{Gluza:2009yy}
with the addition of the ${\mathcal O}(\ep)$ contribution, where $\ep = \frac{(4-D)}{2}$, $D$ being the space-time dimension.
Recently, the calculation of a subset of the three-loop master integrals \cite{Henn:2016kjz} has made it possible 
to obtain the vector form factor 
at three loops \cite{Henn:2016tyf} in the color-planar limit.
While the main goal is to compute the complete three-loop corrections for the form factors, the ${\cal O}(\ep)$ pieces 
at two-loop order are necessary ingredients.
Additionally, computing the master integrals with a different technique to the required order in $\ep$ and cross-checking 
the available results in the literature are also motivating factors. 
In \cite{ours:2017xx}, we compute the contributions to the massive form factors
up to ${\cal O}(\ep^2)$ for different currents, namely, vector, axial-vector, scalar 
and pseudo-scalar currents, which serve as input for ongoing and future 3- and 4-loop calculations.

\section{The heavy quark form factors} \label{sec:theory}

We consider the decay of a virtual massive boson of momentum $q$ into a pair of heavy quarks of mass $m$,
momenta $q_1$ and $q_2$, and color $c$ and $d$,
through a vertex $X_{cd}$, where $X_{cd} = \Gamma^{\mu}_{V,cd}, \Gamma^{\mu}_{A,cd}, 
\Gamma_{S,cd}$ and $\Gamma_{P,cd}$
indicate a vector boson, an axial-vector boson, a scalar and a pseudo-scalar, respectively.
$q^2$ is the center of mass energy squared and we define the dimensionless variable 
% \begin{equation}
$s = \frac{q^2}{m^2}\,.$
% \end{equation}
To eliminate square-roots,
we introduce another dimensionless variable $x$ defined by
\begin{equation}
 s = - \frac{(1-x)^2}{x} \,.
\end{equation}
The amplitudes take the following general form 
\begin{equation}
 \bar{u}_c (q_1) X_{cd} v_d (q_2) \,,
\end{equation}
where $\bar{u}_c (q_1)$ and $v_d (q_2)$ are the bispinors of the quark and the anti-quark, respectively.
We denote the corresponding UV renormalized form factors by $F_{I}$, $I = V, A, S, P$.
They are expanded in the strong coupling constant ($\alpha_s = g_s^2/(4\pi)$) as follows
\begin{equation}
 F_{I} = \sum_{n=0}^{\infty} \asr^n F_{I}^{(n)} \,.
\end{equation}
% 
%-----------------
Studying the general Lorentz structure, one finds the following generic forms for the amplitudes.
For the vector and axial-vector currents  we find
\begin{align}
 \Gamma_{cd}^{\mu} = \Gamma_{V,cd}^{\mu} + \Gamma_{A,cd}^{\mu}
%  \nonumber\\
%  &
 = -i \delta_{cd} & \Big[ v_Q \Big( \gamma^{\mu}~F_{V,1} + \frac{i}{2 m} \sigma^{\mu\nu} q_{\nu} ~F_{V,2} \Big)
\nonumber\\& 
+ a_Q \Big( \gamma^{\mu} \gamma_5~F_{A,1} 
         + \frac{1}{2 m} q^{\mu} \gamma_5 ~ F_{A,2}  \Big) \Big] 
\end{align}
where $\sigma^{\mu\nu} = \frac{i}{2} [\gamma^{\mu},\gamma^{\nu}]$, $q=q_1+q_2$, and $v_Q$ and $a_Q$ are the 
SM vector and axial-vector couplings, respectively.
For the scalar and pseudo-scalar currents, we find
\begin{align}
 \Gamma_{cd} &= \Gamma_{S,cd} +  \Gamma_{P,cd}
%  \nonumber\\
%  &
 = - \frac{m}{v} \delta_{cd} ~ \Big[ s_Q \, F_{S} + i p_Q \gamma_5 \, F_{P} \Big] \,,
\end{align}
where $v = (\sqrt{2} G_F)^{-1/2}$ is the SM Higgs vacuum expectation value, with 
$G_F$ being the Fermi constant, $s_Q$ and $p_Q$ are the scalar and pseudo-scalar couplings, respectively.

To extract the form factors $F_{I,i}, ~ I=V,A$, we multiply the following projectors on $\Gamma^{\mu}_{cd}$ and perform a trace over
the spinor and color indices
\begin{align}
 P_{V,i} &= \frac{i}{v_Q} \frac{\delta_{cd}}{N_c} \frac{\q_2 - m}{m} \Big( \gamma_{\mu} g_{V,i}^{1} + \frac{1}{2 m} (q_{2 \mu} - q_{1 \mu}) g_{V,i}^{2} \Big) \frac{\q_1 + m}{m} \,,
 \nonumber\\
 P_{A,i} &= \frac{i}{a_Q} \frac{\delta_{cd}}{N_c} \frac{\q_2 - m}{m} \Big( \gamma_{\mu} \gamma_5 g_{A,i}^{1} 
          + \frac{1}{2 m} (q_{1 \mu} + q_{2 \mu}) \gamma_5 g_{A,i}^{2} \Big) \frac{\q_1 + m}{m} \,.
\end{align}
$g_{I,i}^{k} \equiv g_{I,i}^{k} (d,s)$ are given in 
\cite{Bernreuther:2004ih, Bernreuther:2004th}\footnote{In \cite{Bernreuther:2004ih}, there is a typo for 
$g_{V,2}^{2}$.
The formula in \cite{Bernreuther:2004th} is correct.}.
$N_c$ denotes the number of colors, 
$C_F = \frac{N_c^2-1}{2 N_c}$ and $C_A = N_c$ are the eigenvalues of the Casimir operators
of the gauge group SU($N_c$) in the fundamental and the adjoint representation, respectively.
% For QCD $N_c=3$ and $T_F = \frac{1}{2}$ is the trace renormalization of the fundamental representation. 
% $n_l$ denotes the number of light quarks.
% 
% 
% where, {\footnote{In \cite{Bernreuther:2004ih}, there is a typo for $g_{V,2}^{2}$.}}
% \begin{align}
%  &g_{V,1}^{1} = - \frac{1}{4 (1-\ep)} \frac{1}{(s-4)} \,,
% %  
% & &g_{V,1}^{2} = \frac{(3-2\ep)}{(1-\ep)} \frac{1}{(s-4)^2} \,,
%  \nonumber\\ 
%  &g_{V,2}^{1} = \frac{1}{(1-\ep)} \frac{1}{s(s-4)} \,,
% %  
% & &g_{V,2}^{2} = - \frac{1}{(1-\ep)} \frac{1}{(s-4)^2} \Bigg( \frac{4}{s} + 2 - 2 \ep \Bigg) \,,
%  \nonumber\\ 
%  &g_{A,1}^{1} = - \frac{1}{4 (1-\ep)} \frac{1}{(s-4)} \,,
% %  
% & &g_{A,1}^{2} = - \frac{1}{(1-\ep)} \frac{1}{s(s-4)} \,,
%  \nonumber\\ 
%  &g_{A,2}^{1} = \frac{1}{(1-\ep)} \frac{1}{s(s-4)} \,,
% %  
% & &g_{A,2}^{2} = \frac{1}{(1-\ep)} \frac{1}{s^2 (s-4)} \Bigg( 4 (3-2 \ep) - 2 s (1-\ep) \Bigg) \,.
% \end{align}
% % 
% 
% 
The form factors $F_S$ and $F_P$ can be obtained from $\Gamma_{cd}$ through suitable projectors as given below
and performing trace over the spinor and color indices 
\begin{align}
 P_{S} &= \frac{v}{2 m s_Q} \frac{\delta_{cd}}{N_c} \frac{\q_2 - m}{m} \Bigg( - \frac{1}{(s-4)} \Bigg) \frac{\q_1 + m}{m} \,,
 \nonumber\\
 P_{P} &= \frac{v}{2 m p_Q} \frac{\delta_{cd}}{N_c} \frac{\q_2 - m}{m} \Bigg( - \frac{i}{s} \gamma_5 \Bigg) \frac{\q_1 + m}{m} \,.
\end{align}

% 
% %%%%%%%%%%%%%%%%%%%%%%%%%%%%%%%%%%%
% 

\subsection{Renormalization}
To regularize the unrenormalized form factors 
we use dimensional regularization \cite{tHooft:1972tcz} in $D=4-2\ep$ space-time dimensions. 
To do so, it becomes important to define $\gamma_5$ in a proper manner within 
this regularization scheme. Based on the appearance of $\gamma_5$ in a $\gamma$-chain
in the axial-vector and pseudo-scalar form factors, the Feynman diagrams can be subdivided 
into two categories: non-singlet contributions, where $\gamma_5$ is attached to open fermion 
lines and singlet contributions, where $\gamma_5$ is attached to a closed fermion loop.
For the non-singlet case, we use an anticommuting $\gamma_5$ in $D$ space-time dimensions with $\gamma_5^2 = 1$,
as it does not lead to any spurious singularities.
In this case, a canonical Ward identity holds to this order, as described by Eq.~\eqref{eq:cwi}.
We follow the prescription presented in 
\cite{Akyeampong:1973xi,Larin:1993tq}, which mostly followed \cite{tHooft:1972tcz},
for the $\gamma_5$'s in the singlet contributions. 
For each $\gamma_5$ in a fermion loop we use
\begin{equation}
 \gamma_5 = \frac{i}{4!} \epsilon_{\mu\nu\rho\sigma} \gamma^{\mu} \gamma^{\nu} \gamma^{\rho} \gamma^{\sigma},
\end{equation}
where the Lorentz indices are $D$-dimensional. In the end, we are left with  
the product of two $\epsilon$-tensors which is expressed 
in terms of $D$-dimensional metric tensors. 
This prescription of $\gamma_5$ needs a special treatment during renormalization, as will be discussed 
later.

The ultraviolet (UV) renormalization is performed in a mixed scheme.
We renormalize the heavy-quark mass and wave function in the on-shell (OS) 
scheme, while the strong coupling constant is renormalized 
in the modified minimal subtraction ($\overline{\rm MS}$) scheme \cite{tHooft:1973mfk, Bardeen:1978yd}.
The corresponding renormalization constants are already known in the literature and are denoted by 
$Z_{m, {\rm OS}}$ \cite{Broadhurst:1991fy, Melnikov:2000zc,Marquard:2007uj}, 
$Z_{2,{\rm OS}}$ \cite{Broadhurst:1991fy, Melnikov:2000zc,Marquard:2007uj} and 
$Z_{a_s}$ \cite{Gross:1973id, Politzer:1973fx, Caswell:1974gg, Jones:1974mm, Egorian:1978zx}
for the heavy-quark mass, wave function and strong coupling constant, respectively. 
The renormalization of massive fermion lines has been taken care of by properly 
considering the counterterms. The singlet contributions demand extra care for renormalization.
The singlet pieces of the axial-vector current are infrared (IR) finite but
the chirality preserving part of them
contains a UV pole
which is renormalized by the multiplicative renormalization constant $Z_J$. 
Larin's prescription \cite{Larin:1993tq} for $\gamma_5$, on the other hand, implies multiplication 
of a finite renormalization constant $Z_5^{fin}$ which ensures that the anomalous
Ward identity Eq.~\eqref{eq:awi}, as shown below, is satisfied.
We would like to note that the Ward identities are true for physical quantities and 
hence, the remaining finite renormalization due to $Z_5^{fin}$ has to be carried out 
in calculating finally the corresponding observable to which the corresponding form factor contributes.
An additional heavy quark mass renormalization is needed for scalar and pseudo-scalar currents 
due to the presence of heavy-quark mass in the Yukawa coupling. 
The singlet piece of the pseudo-scalar vertex is both IR and UV finite, hence 
no additional renormalization is necessary.

\subsection{Infrared structure}

The IR singularities of the massive form factors can be factorized \cite{Becher:2009kw} as a multiplicative 
renormalization factor. The corresponding structure is constrained by the renormalization group equation (RGE),
\begin{equation}
 F_{I} = Z (\mu) F_{I}^{fin} (\mu)
\end{equation}
where $F_{I}^{fin}$ is finite as $\ep \rightarrow 0$ and the RGE of $Z$ gives
\begin{equation} \label{eq:rgeZ}
 \frac{d}{d \ln \mu} \ln Z(\ep, x, m, \mu)  = - \Gamma (x,m,\mu) \,.
\end{equation}
Note that $Z$ does not carry any information ($I$) regarding the vertex. 
Here $\Gamma$ denotes the massive cusp anomalous dimension, which is available
up to three-loop level \cite{Korchemsky:1987wg,Korchemsky:1991zp,Grozin:2014hna,Grozin:2015kna}.
Both $Z$ and $\Gamma$ can be expanded in a perturbative series in $\alpha_s$
\begin{equation}
 Z = \sum_{n=0}^{\infty} \asr^n Z^{(n)} \,, \qquad
 \Gamma = \sum_{n=0}^{\infty} \asr^{n+1} \Gamma_{n}
\end{equation}
and the solution for Eq.~\eqref{eq:rgeZ} is given by
\begin{equation} \label{eq:solnZ}
 Z = 1 + \asr \Bigg[ \frac{\Gamma_0}{2 \ep} \Bigg] 
   + \asr^2 \Bigg[ \frac{1}{\ep^2} \Big( \frac{\Gamma_0^2}{8} - \frac{\beta_0 \Gamma_0}{4} \Big) + \frac{\Gamma_1}{4 \ep} \Bigg] 
   + {\cal O} (\alpha_s^3) \,.
\end{equation}
Eq.~\eqref{eq:solnZ} correctly predicts the IR singularities for all massive form factors at two-loop level. 
% 
%%%%%%%%%%%%%%%%%%%%%%%%%%%%%%%%%%%%%

%%%%%%%%%%%%%%%%%%%%%%%%%%%%%%%%%%%
\subsection{Anomaly and Ward identities}
As stated earlier, the axial-vector and pseudo-scalar currents consist of two 
different contributions: non-singlet and singlet, depending on whether the vertex 
is attached to open fermion lines or a fermion loop, as
\begin{equation}
 \Gamma_{A,cd}^{\mu} = \Gamma_{A,cd}^{\mu, ns} + \Gamma_{A,cd}^{\mu, s} \,, \quad
 \Gamma_{P,cd} = \Gamma_{P,cd}^{ns} + \Gamma_{P,cd}^{s} \,.
\end{equation}
$ns$ and $s$ denote non-singlet and singlet cases, respectively.
For the non-singlet case, 
we use anti-commutation of $\gamma_5$ and finally $\gamma_5^2 = 1$. This approach 
respects the chiral invariance and leaves us with the following Ward identity 
\begin{equation} \label{eq:cwi}
 q^{\mu} \Gamma_{A,cd}^{\mu, ns} = 2 m \Gamma_{P,cd}^{ns} \,.
\end{equation}
% which in terms of the form factors, takes the following form
% \begin{equation} \label{eq:cwiFF}
%  2 F_{A,1}^{ns} + \frac{F_{A,2}^{ns}}{2} \Big( - \frac{(1-x)^2}{x} \Big) = 2 m F_{P}^{ns} \,.
% \end{equation}
% 
The singlet contributions exhibit the ABJ anomaly \cite{Adler:1969gk, Bell:1969ts} which involves the truncated matrix 
element of the gluonic operator $G\tilde{G}$ between the vacuum and a pair of heavy quark states.
% The gluonic operator is described by
% \begin{equation}
%  G(x) \tilde{G} (x) \equiv \varepsilon_{\mu\nu\rho\sigma} G^{a,\mu\nu} (x) G^{a,\rho\sigma} \,,
% \end{equation}
% where $G^{a,\mu\nu}$ represents gluonic field strength tensor.
Denoting its contribution by $\langle G\tilde{G} \rangle_Q$, we can immediately write down the anomalous Ward 
identity for the singlet case, as follows
\begin{equation} \label{eq:awi}
 q_{\mu} \Gamma_{A,cd}^{\mu, s} = 2 m \Gamma_{P,cd}^{s} - i \asr T_F \langle G\tilde{G} \rangle_Q \,.
\end{equation}
% which implies 
% \begin{equation}
%  2 F_{A,1}^{s} + \frac{F_{A,2}^{s}}{2} \Big( - \frac{(1-x)^2}{x} \Big) = 2 m F_{P}^{s} - i \asr T_F F_{G,Q} \,.
% \end{equation}
%-------------
The UV renormalization of the quantity $\langle G\tilde{G} \rangle_Q$ involves mixing of the gluonic operator 
with another operator $\partial_{\mu} \bar{\psi} \gamma^{\mu} \gamma_5 \psi$, as discussed in 
\cite{Larin:1993tq, Zoller:2013ixa, Ahmed:2015qpa}.
%------------
% UV renormalization of the quantity $F_{G,Q}$ involves mixing of the gluonic operator $G\tilde{G}$ 
% with another operator $\partial_{\mu} \bar{\psi} \gamma^{\mu} \gamma_5 \psi$, as discussed in 
% \cite{Larin:1993tq, Zoller:2013ixa, Ahmed:2015qpa}, where $\psi$ indicates all quark flavors including 
% the massive one, as 
% \begin{equation}
%  F_{G,Q} = Z_{GG} \hat{F}_{G,Q} + Z_{GJ} \hat{F}_{J,Q} \,.
% \end{equation}
% $\hat{F}_{J,Q}$ indicates bare contribution from the second operator and, $Z_{GG}$ \& $Z_{GJ}$
% are the corresponding renormalization constant.

% For the non-singlet case, 
% \begin{equation}
%  q_{\mu} \Gamma^{\mu}_5 = 2 m \Gamma_5
% \end{equation}
% which implies 
% \begin{equation}
%  2 F_{A,1}^{R} + \frac{F_{A,2}^{R}}{2} \Big( - \frac{(1-x)^2}{x} \Big) = 2 F_{P}^{R}
% \end{equation}
% 
% where $Z_{\rm fin}^{(2)}$ can be obtained demanding that ABJ anomaly is maintained
% \begin{equation}
%  F_{A,i}^{s,(2)} = 2 i m F_{PS,i}^{s,(2)} + T_F F_{G,i}^{(1)}
% \end{equation}
% where
% \begin{equation}
%  F_{G,i}^{(1)} = Z_{GG} \hat{F}_{G,i}^{(1)} + Z_{GJ} \hat{F}_{J,i}^{(1)}
% \end{equation}

\section{Details of the computation}

The computation of the two-loop form factors has been performed following the generic procedure.
We have used {\tt QGRAF} \cite{Nogueira:1991ex} to generate 
the Feynman diagrams. The output has then been processed 
using {\tt FORM} \cite{Vermaseren:2000nd, Tentyukov:2007mu}
to perform the Lorentz, Dirac and color algebra. Specifically we use the {\tt FORM} package
{\tt color} \cite{vanRitbergen:1998pn} for color algebra. The diagrams have been expressed 
in terms of a linear combination of a large set of scalar integrals.
These integrals have been reduced to a set of master integrals (MIs) using 
integration by parts identities 
(IBPs) \cite{Lagrange:IBP, Gauss:IBP, Green:IBP, Ostrogradski:IBP, Chetyrkin:1980pr, Chetyrkin:1981qh, Tkachov:1981wb} 
with the help of the program {\tt Crusher} \cite{Crusher}.
The diagrams have been matched to the different topologies 
defined in {\tt Crusher} using the codes {\tt Q2e/Exp} \cite{Harlander:1997zb,Seidensticker:1999bb}.
Now, after performing the reductions, all that remains to be done is to compute the MIs. 
% In the following sections, we present the methods we used to achieve this.
We follow both the method of differential equations and the method of difference equations
to achieve this.

\subsection{Method of differential equations}

We have obtained the two-loop MIs contributing to massive form factors as Laurent expansions in $\varepsilon$ 
by means of the standard differential equation method \cite{Kotikov:1990kg,Kotikov:1991hm,Kotikov:1991pm,Remiddi:1997ny,Kotikov:2010gf,Henn:2013pwa}. 
This technique has already been applied to such type of integrals at two and three loops in 
\cite{Bonciani:2003te,Bonciani:2003hc,Henn:2016kjz}. 
In this work, we have calculated the two-loop MIs up to sufficient order in $\varepsilon$ to obtain
$O(\varepsilon^{2})$ accuracy in the form factors. 
%
%-----------------------------------------
% % 
% The MIs arising in non-singlet contributions can be expressed in terms of the following single integral family
% \begin{equation}
% J(\nu_1,\dots,\nu_7) = \Big((4\pi)^{2-\varepsilon}e^{\varepsilon\gamma_E}\Big)^2\,\int\frac{d^Dl_1d^Dl_2}{(2\pi)^{2D}}\,\frac{1}{D_1^{\nu_1} \dots D_7^{\nu_7}}, 
% \end{equation}
% where 
% \begin{eqnarray}
% && D_1 = (l_1+q_1)^2-m^2, \quad D_2 = (l_2+q_1)^2-m^2, \quad D_3 = (l_1-q_2)^2-m^2, \nonumber \\
% && D_4 = (l_2-q_2)^2-m^2, \quad D_5 =l_1^2, \quad D_6 = (l_1-l_2)^2, \quad D_7 = (l_1-l_2+q_2)^2-m^2.
% \end{eqnarray}
% % 
% In the case of singlet contributions, the MIs are given by
% \begin{equation}
% K(\nu_1,\dots,\nu_6) = \Big((4\pi)^{2-\varepsilon}e^{\varepsilon\gamma_E)}\Big)^2\,\int\frac{d^Dl_1d^Dl_2}{(2\pi)^{2D}}\,\frac{1}{D_1^{'\,\nu_1} \dots D_6^{'\,\nu_6}}, 
% \end{equation}
% where
% \begin{eqnarray}
% && D_1' = (l_1+q_1)^2, \quad D_2' = (l_2+q_1)^2-m^2, \quad D_3' = (l_1-q_2)^2, \nonumber \\
% && D_4' = (l_2-q_2)^2-m^2, \quad D_5' =l_1^2, \quad D_6' = (l_1-l_2)^2-m^2 \,.
% \end{eqnarray}
% % 
%-------------------------------------
% For both cases, 
We have derived a system of coupled linear differential equations by taking derivative of each MI \textit{w.r.t} $x$
and then using IBPs again with help of {\tt{Crusher}}. The system can then be expanded 
order-by-order in the parameter $\ep$.
The expanded system simplifies greatly and can be
arranged mostly in a block triangular form except for a few $2\times2$
sub-systems, for which we first decouple them and use the variation of constants to solve.
Generically, we solve the whole system in a bottom-up approach \textit{i.e.} first solving the simplest sectors and then 
moving up in the chain of sub-systems. 
These steps have been automated and results have been obtained efficiently using 
a minimal set of independent harmonic polylogarithms (HPLs) 
by means of the \verb+Mathematica+ packages \verb+Sigma+ \cite{Schneider:sigma1,Schneider:sigma2} 
and \verb+HarmonicSums+ \cite{Ablinger:2014rba, Ablinger:2010kw, Ablinger:2013hcp, Ablinger:2011te, Ablinger:2013cf, Ablinger:2014bra}. 

Now what remains is to obtain the appropriate boundary conditions. As noticed earlier in \cite{Bonciani:2003te, Bonciani:2003hc}, 
the analytic structure of the MIs puts strong constraints on the choice of integration constants. 
For most of the MIs, we determine the boundary conditions by demanding regularity of the functions at $x=1$.
However, some MIs are characterized by a branch cut at $x=1$ and for such cases, we have matched the general solutions of the differential equations with 
asymptotic expansions of the corresponding integrals around $x\rightarrow 1$.

\subsection{Method of difference equations}
We have considered the method of difference equation as an alternative way to compute the MIs. 
The idea \cite{Remiddi:1997ny} is to write the integrals in series expansion of $y=(1-x)$
and then use the differential equations to derive difference equations satisfied by the coefficients of the series.
In the non-singlet case, considering the fact that the MIs are regular at $x=1$, we can 
therefore write
\begin{equation}
J_i(y)=\sum_{n=0}^{\infty} \sum_{j=-2}^r \varepsilon^j C_{i,j}(n) y^n\,.
\label{NSMIyexp}
\end{equation}
On the other hand, for the singlet case, some integrals have a branch cut at $x=1$ which actually shows up as $\ln(1-x)\equiv\ln(y)$. 
Henceforth, we include powers of these logarithms in the expansion of these integrals as \cite{Maier:2007yn}
\begin{equation}
K_i(y)=\sum_{n=0}^{\infty} \sum_{k=0}^3 \sum_{j=-2}^r \varepsilon^j C_{i,j,k}(n) \ln^k(y) y^n.
\label{SMIyexp}
\end{equation}
Using the system of differential equations, we have obtained a system of difference equations 
for the coefficients $C_{i,j}$ and $C_{i,j,k}$. This system now can be solved
with proper initial conditions in the same manner as for the system of differential equations
and finally we have obtained the MIs in terms of harmonic sums and generalized harmonic sums and after 
performing the sums, in terms of HPLs. The whole procedure has been automated using
\verb+Sigma+,
\verb+EvaluateMultiSums+,
\verb+SumProduction+ \cite{EMSSP}
and \verb+HarmonicSums+.

\section{Results}

The analytic results for all two-loop UV renormalized form factors $F_I$, $I=V,A,S,P$ up to ${\cal O}(\ep^2)$
are presented as an attachment \verb|anci.m| with the arXiv preprint \cite{ours:2017xx}. 
The results up to ${\cal O}(\ep)$ are printed in the appendix of \cite{ours:2017xx}.
% 
% As an interesting fact, around 300 independent HPLs appear 
% In the two-loop results, the following two constants appear
% \begin{align}
%  {c}_1 &= 12 \zeta_2 \ln^2(2) + \ln^4(2) + 24 {\rm Li}_4 \Big( \frac{1}{2} \Big)
% \nonumber\\
%  c_2 &= 26 \zeta_2^2 \ln (2) - 20 \zeta_2 \ln^3(2) - \ln^5(2) + 120 {\rm Li}_5 \Big( \frac{1}{2} \Big) \,.
% \end{align}
% 

\begin{figure}[htb]
\centerline{%
\includegraphics[width=0.45\textwidth]{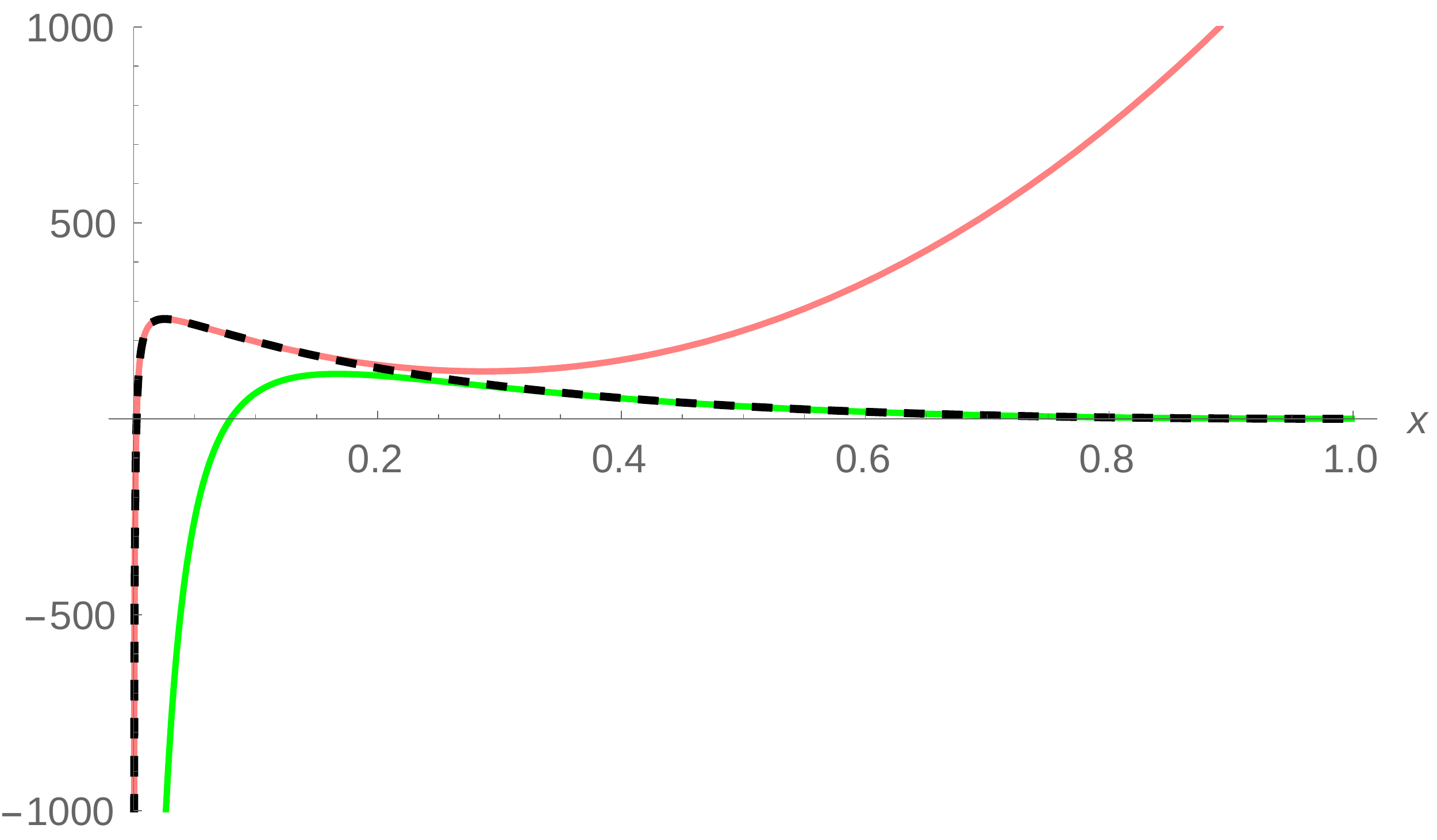}
\includegraphics[width=0.45\textwidth]{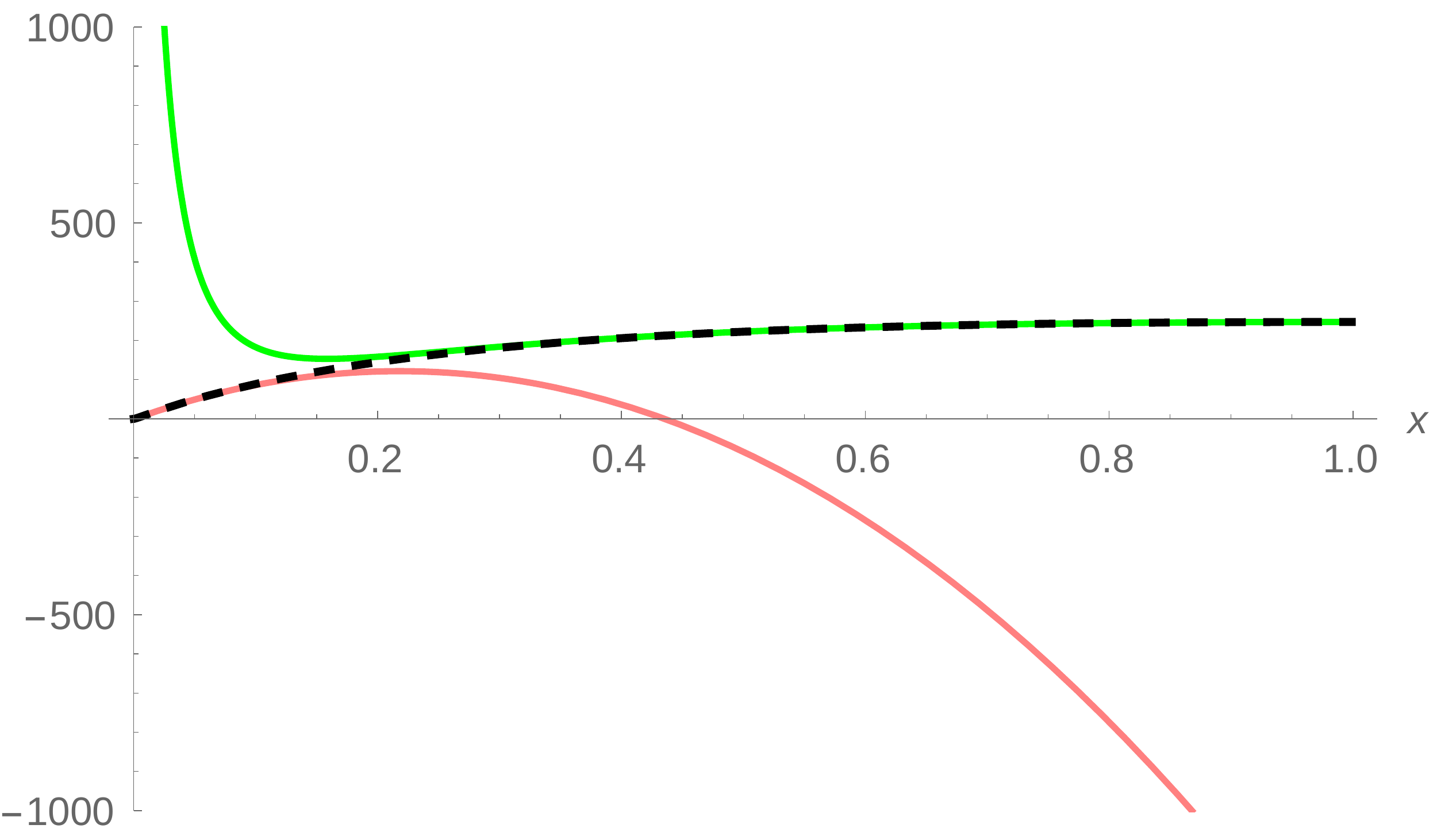}}
\caption{$C_A C_F$ coefficient of ${\cal O}(\ep)$ part of $F_{V,1}$ (left) and $F_{V,2}$ (right).}
\label{Fig:VF12cacfep1}
\end{figure}

\begin{figure}[htb]
\centerline{%
\includegraphics[width=0.45\textwidth]{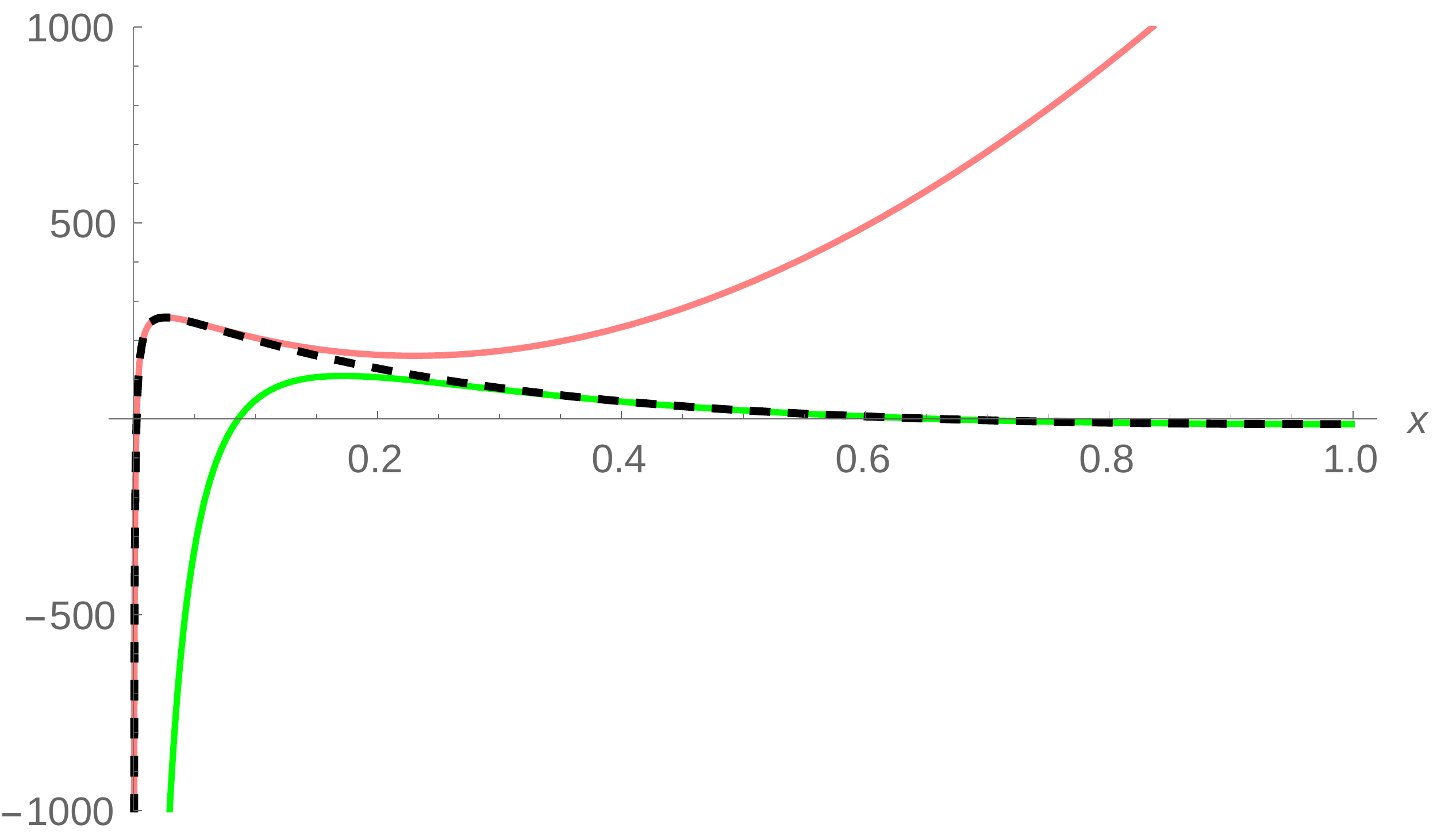}
\includegraphics[width=0.45\textwidth]{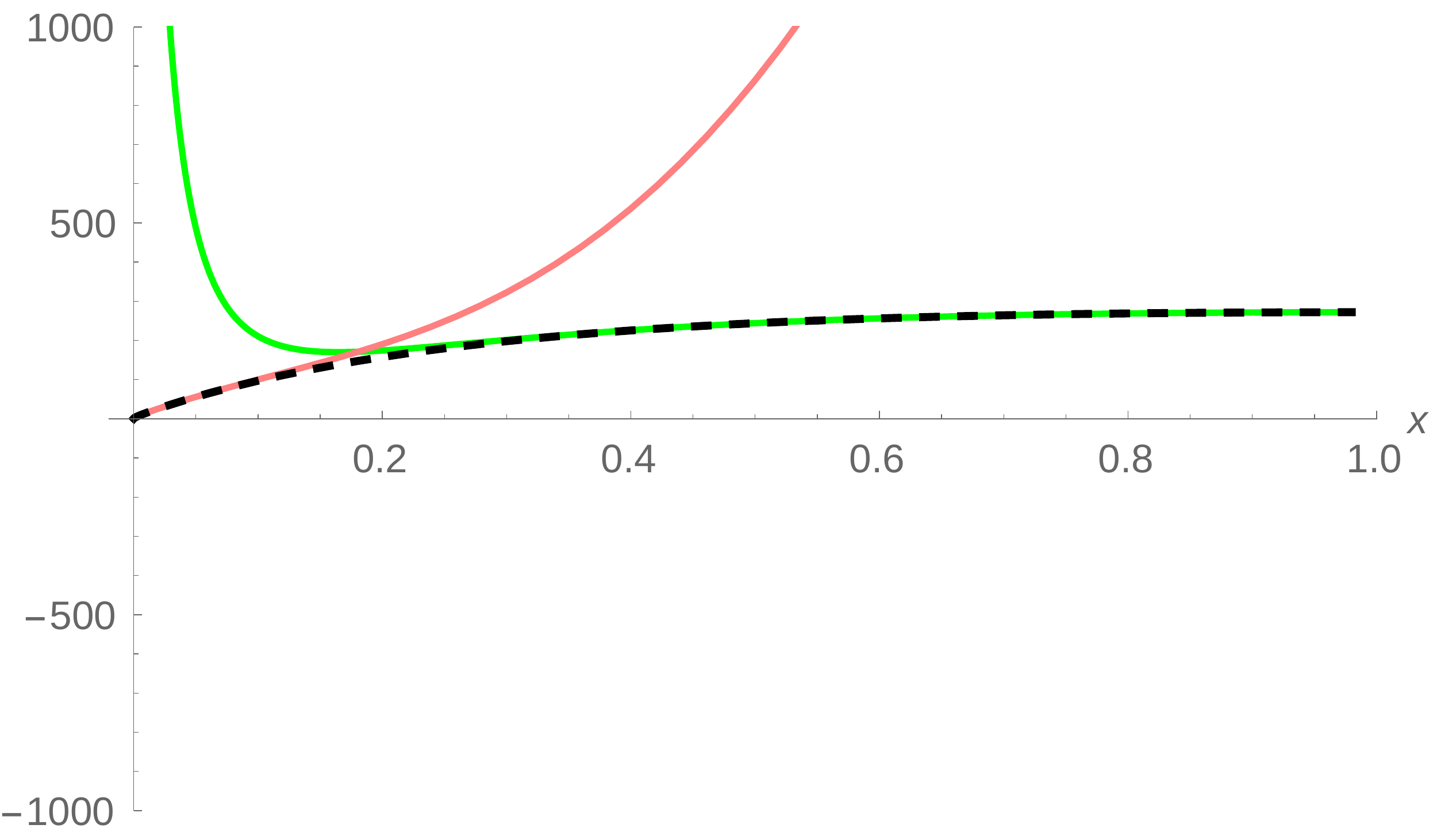}}
\caption{$C_A C_F$ coefficient of ${\cal O}(\ep)$ part of $F_{A,1}$ (left) and $F_{A,2}$ (right).}
\label{Fig:AVG12cacfep1}
\end{figure}

\begin{figure}[htb]
\centerline{%
\includegraphics[width=0.45\textwidth]{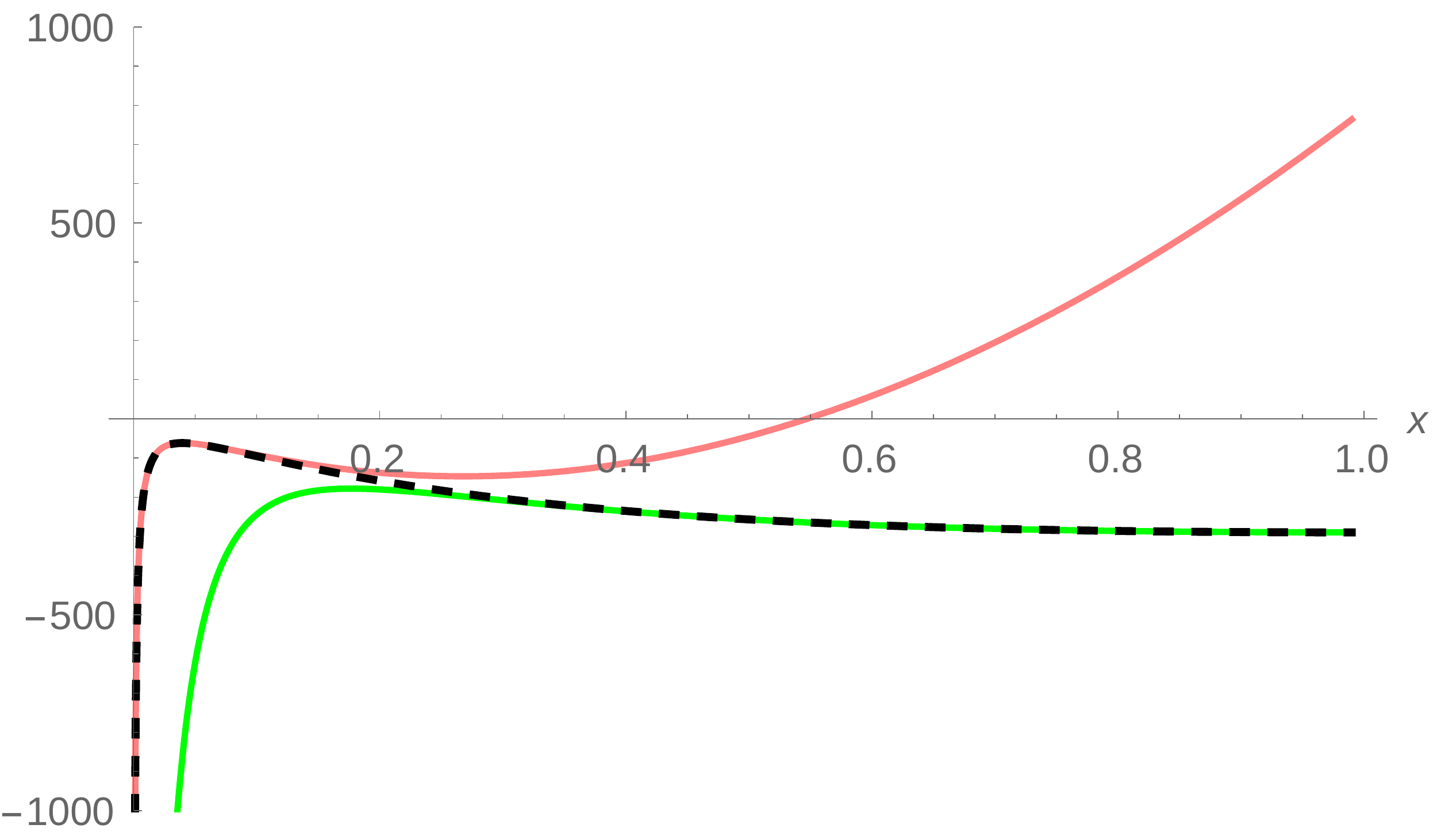}
\includegraphics[width=0.45\textwidth]{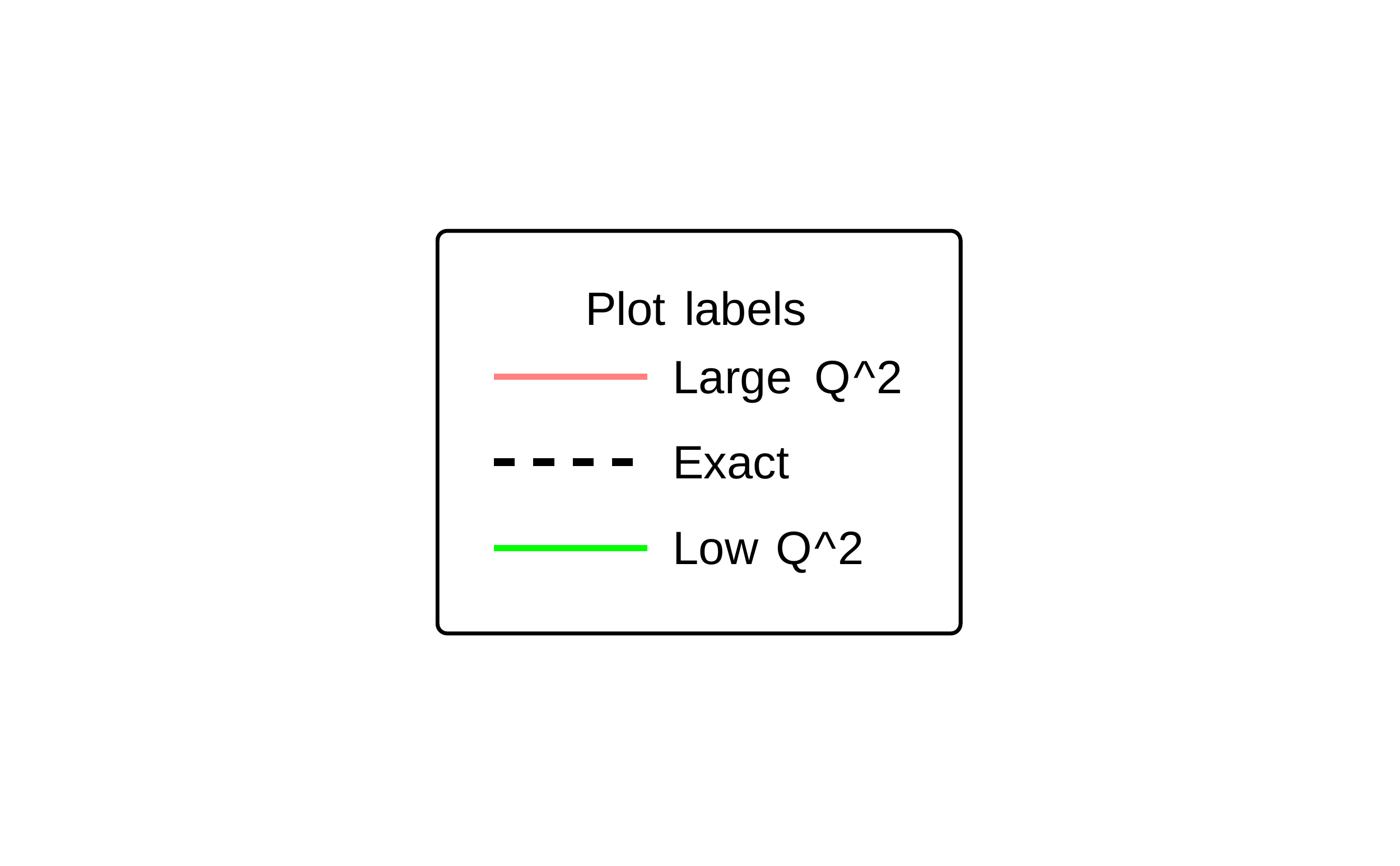}}
\caption{$C_A C_F$ coefficient of ${\cal O}(\ep)$ part of $F_{S}$ (left) and labels for all plots.}
\label{Fig:Scacfep1}
\end{figure}

The behavior of the form factors in various kinematic regions also carries substantial importance.
We therefore study them in 
the low energy, high energy and threshold regions which correspond to $x \rightarrow 1$, 
$x \rightarrow 0$ and $x \rightarrow -1$, respectively.
We extensively use the packages \verb|Sigma| and \verb|HarmonicSums|
% \verb|Sigma| \cite{Schneider:sigma1,Schneider:sigma2} and 
% \verb|HarmonicSums| \cite{Ablinger:2014rba, Ablinger:2010kw, Ablinger:2013hcp, Ablinger:2011te, Ablinger:2013cf, Ablinger:2014bra}
for all the expansions.
Below, we present a brief summary of all the expansions, and instead of printing the voluminous results 
here, we 
choose to plot some parts of them, namely the coefficient of $C_A C_F$ for the ${\cal O} (\ep)$ piece of each of the form factors 
and the corresponding expansion in high and low energy regions. Fig.~\ref{Fig:VF12cacfep1}, Fig.~\ref{Fig:AVG12cacfep1} and Fig.~\ref{Fig:Scacfep1}
contain the corresponding terms for vector, axial-vector and scalar form factors, respectively.
The notation for the figures is presented in the right part of Fig.~\ref{Fig:Scacfep1}.

\emph{Low energy region} ($q^2 \ll 4 m^2$): The low energy limit of the space-like ($q^2 < 0$)
form factors is given by $x\rightarrow1$. 
To expand the HPLs, we redefine $x$ as $x=e^{i\phi}$ and expand them around $\phi=0$.
Note that for $\phi\rightarrow0$, $F_{V,1}=1$ and $F_{V,2}$ is finite and agrees with the anomalous magnetic moment 
of the top quark, as expected.

\emph{High energy region} ($q^2 \gg 4 m^2$): The asymptotic or high energy limit is given by $x\rightarrow0$. We expand the form 
factors up to ${\cal O} (x^4)$. In the limit $x\rightarrow0$, the chirality flipping form factors $F_{V,2}$ and $F_{A,2}$ vanish
and the effect of $\gamma_5$ gets nullified implying $F_{V,1}=F_{A,1}$ and $F_S=F_P$.

\emph{Threshold region} ($q^2 \sim 4 m^2$): In the threshold limit $q^2 \sim 4 m^2$ or $x\rightarrow-1$, we define the variable
% \begin{equation}
$\beta = \sqrt{1 - \frac{4m^2}{q^2}}$
% \end{equation}
and expand the form factors around $\beta=0$ up to ${\cal O}(\beta^2)$.

\subsection{Checks}
We explicitly check our results by comparing them to the ones available in the literature.
Except a difference in an overall factor due to different renormalization schemes and another difference in the wave 
function renormalization ($Z_{2,{\rm OS}}$), we agree 
with both the bare and UV renormalized results in \cite{Bernreuther:2004ih, Bernreuther:2004th, Bernreuther:2005gw} up to ${\cal O}(\ep^0)$
for all the form factors except the singlet parts of axial-vector form factors.
While the bare singlet contributions for axial-vector currents also match with the results in \cite{Bernreuther:2005rw},
we find a mismatch of terms which are polynomial in $x$ for the renormalized contributions.
We also compare the ${\cal O}(\ep)$ pieces for the two-loop vector form factors $F_{V,1}^{(2)}$ and $F_{V,2}^{(2)}$ with the results 
presented in \cite{Gluza:2009yy} and find a difference of the following term, as has also been mentioned in 
\cite{Henn:2016tyf},
\begin{equation}
 - C_F C_A \Bigg[ \ep \Big\{ \frac{1037 x^3}{(1+x)^6} \Big\} \Bigg] \,.
\end{equation}
We cross-checked the vector form factors, the exact ones and also their expansions in different regions, 
presented in \cite{Henn:2016tyf} up to 
${\cal O} (\ep)$ in the color-planar limit.
We also compare with predictions of the vector form factor $F_{V,1}$ in the high energy limit as given 
in \cite{Gluza:2009yy} considering the evolution equations.

\section{Conclusion}

To shed more light on the Higgs mechanism and electro-weak symmetry breaking, a precise determination of the properties 
of the top quark, the heaviest SM particle, is needed. A future electron-positron collider can reach 
high precision and hence an equal theory prediction is much required. In a similar way this also applies
to the LHC for its high luminosity phase. 
In \cite{ours:2017xx}, we compute up to ${\cal O}(\ep^2)$ contributions to the heavy quark form factors
for vector, axial-vector, scalar and pseudo-scalar currents at two-loop level. 
These contributions constitute an important part in three-loop results and also contribute to potential
future 4-loop calculations.
Additionally, they serve as a cross-check of earlier results available in the literature.

%uncomment the following lines to place a figure
%\begin{figure}[htb]
%\centerline{%
%\includegraphics[width=12.5cm]{Fig1}}
%\caption{Plot of ...}
%\label{Fig:F2H}
%\end{figure}

%%%%%%%%%%%%%%%%%%%%%%%%
\bibliography{main}
\bibliographystyle{MYutphysM}

\end{document}